\documentclass[sigconf,screen]{acmart}

\usepackage{enumitem}

\makeatletter
\def\@copyrightpermission{
  \includegraphics[height=5ex]{doclicense-CC-\ACM@cc@type-88x31}\\
  This work is licensed under a \href{https://creativecommons.org/licenses/by/4.0/}{Creative Commons Attribution International\\
  4.0 License}.
  \vspace{4mm}}
\makeatother

\copyrightyear{2024}
\acmYear{2024}
\setcopyright{rightsretained}
\acmConference[CSET 2024]{Workshop on Cyber Security Experimentation and Test}{August 13, 2024}{Philadelphia, PA, USA}
\acmBooktitle{Workshop on Cyber Security Experimentation and Test (CSET 2024), August 13, 2024, Philadelphia, PA, USA}
\acmDOI{10.1145/3675741.3675754}
\acmISBN{979-8-4007-0957-9/24/08}

\begin{document}

\title{Introducing a Comprehensive, Continuous, and Collaborative Survey of Intrusion Detection Datasets}

\author{Philipp Bönninghausen}
\email{philipp.boenninghausen@fkie.fraunhofer.de}
\orcid{0009-0004-7870-7388}
\author{Rafael Uetz}
\email{rafael.uetz@fkie.fraunhofer.de}
\orcid{0000-0002-1592-4607}
\affiliation{%
  \institution{Fraunhofer FKIE}
  \city{Wachtberg}
  \country{Germany}
}

\author{Martin Henze}
\email{henze@spice.rwth-aachen.de}
\orcid{0000-0001-8717-2523}
\affiliation{%
  \institution{RWTH Aachen University}
  \city{Aachen}
  \country{Germany}
}
\affiliation{%
  \institution{Fraunhofer FKIE}
  \city{Wachtberg}
  \country{Germany}
}

\begin{abstract}

Researchers in the highly active field of intrusion detection largely rely on public datasets for their experimental evaluations.
However, the large number of existing datasets, the discovery of previously unknown flaws therein, and the frequent publication of new datasets make it hard to select suitable options and sufficiently understand their respective limitations.
Hence, there is a great risk of drawing invalid conclusions from experimental results with respect to detection performance of novel methods in the real world.
While there exist various surveys on intrusion detection datasets, they have deficiencies in providing researchers with a profound decision basis since they lack comprehensiveness, actionable details, and up-to-dateness.
In this paper, we present \textsc{Comidds}, an ongoing effort to comprehensively survey intrusion detection datasets with an unprecedented level of detail, implemented as a website backed by a public GitHub repository.
\textsc{Comidds} allows researchers to quickly identify suitable datasets depending on their requirements and provides structured and critical information on each dataset, including actual data samples and links to relevant publications.
\textsc{Comidds} is freely accessible, regularly updated, and open to contributions.
  
\end{abstract}

\ccsdesc[500]{Security and privacy~Intrusion/anomaly detection and malware mitigation}
\ccsdesc[100]{Applied computing~Enterprise computing}
\ccsdesc[100]{Computing methodologies~Modeling and simulation}

\keywords{Intrusion Detection, Dataset, Log Data, Netflow Data, Cyberattack, Enterprise Network, Testbed, Cyber Range, Simulation, Survey}

\maketitle

\section{Introduction}
\label{sec:introduction}

Intrusions of enterprise networks continue to affect thousands of organizations each year, often resulting in data theft, sabotage, and extortion~\cite{verizon2023data}.
Detecting such intrusions in a timely manner is difficult~\cite{axelsson2000base,sommer2010outside}, yet crucial to stop adversaries from reaching their final goals~\cite{lord2022investments}.
It is thus hardly surprising that intrusion detection is a highly active area of research for more than three decades now, with thousands of papers being published each year~\cite{kumar2021nidstrends}.

A large number of these works propose novel intrusion detection methods~\cite{khraisat2019survey,yang2022anomalysurvey,bridges2019hidssurvey} and consequently require realistic data (resembling both benign and adversarial activity) to evaluate them against.
Since many researchers lack access to enterprise networks or permission to run representative attacks against them, there is a high demand for appropriate public datasets~\cite{kenyon2020fit}.
In addition, public datasets (in contrast to private ones) allow for quantitative comparisons of works by different authors as well as independent analyses of the dataset itself to discover potential flaws~\cite{uetz2021socbed}.

Reacting to this high demand, researchers have created a multitude of datasets, which vary greatly in objective, age, and effort put into them~\cite{kenyon2020fit}.
Their contents cover a wide range of environments (e.g., office, cloud, or industrial context), activity (e.g., real or simulated benign activity as well as various attacks), and data formats (e.g., network flows, host log files, or system call traces)~\cite{ring2019survey,bridges2019hidssurvey}.

Since there is no central registry for such datasets and relevant publications are spread over a large number of media and years, researchers may struggle to find datasets fitting their requirements and to fully understand their limitations and potential deficiencies~\cite{kenyon2020fit}.
In particular, some of the most popular and widely used datasets~\cite{kumar2021nidstrends} show significant weaknesses~\cite{kenyon2020fit,tavallaee2009kdd,engelen2022pillars,lanvin2023errors,mchugh2000testing}.
Consequently, researchers using datasets should have an adequate knowledge of available datasets and their characteristics to avoid drawing invalid conclusions from experimental results.

To spare researchers from having to read hundreds of papers before using a dataset, various surveys give an overview of available datasets as well as independent analyses thereof (cf. Section~\ref{sec:relatedwork}).
However, they suffer from three fundamental shortcomings: (1)~They are \emph{static} in the sense that they cannot be updated or corrected once published, (2)~their descriptions of datasets are mostly superficial due to limited space, and (3)~contained data (such as tables and plots) cannot be sorted, filtered, or otherwise processed automatically, e.g., to narrow down choices or create own statistics.

Addressing these shortcomings, we present \textsc{Comidds} -- a \textbf{com}pre\-hensive, continuous, and collaborative \textbf{i}ntrusion \textbf{d}etection \textbf{d}atasets \textbf{s}urvey.
\textsc{Comidds} is freely accessible as a website backed by a public GitHub repository~\cite{idd}, thus allowing for ongoing extensions, corrections, and change tracking.
It provides an overview of key characteristics of all surveyed datasets (currently~48) and dedicated pages for each dataset containing detailed, structured, and critical information on their environment, activity, data format, related publications, and exemplary data snippets.
Thus, \textsc{Comidds} assists researchers in finding and selecting appropriate datasets for their experiments and furthermore raises awareness of known limitations, eventually fostering advances in real-world intrusion detection.
Overall, we make the following contributions:
\begin{itemize}[leftmargin=*]
  \item We introduce \textsc{Comidds}~\cite{idd}, a novel effort to survey intrusion detection datasets based on a GitHub repository (Section~\ref{sec:survey}).
  \item We describe our methodology for finding relevant datasets, reviewing them, and adding them to \textsc{Comidds} (Section~\ref{sec:methodology}).
  \item We visualize key characteristics of the datasets surveyed so far to showcase our machine-readable survey data (Section~\ref{sec:statistics}).
  \item We compare \textsc{Comidds} to existing surveys, showing that it overcomes all shortcomings that we identified (Section~\ref{sec:relatedwork}).
\end{itemize}

\section{\textsc{Comidds}: A Repository-Based Survey of Intrusion Detection Datasets}
\label{sec:survey}

We begin with giving an overview of \textsc{Comidds}~\cite{idd}, including its goals, scope, and current features in the following.
Based on this, we describe our methodology for finding, reviewing, and adding datasets to \textsc{Comidds} in Section~\ref{sec:methodology}.

\textsc{Comidds}' purpose is to aid researchers in finding and selecting suitable datasets to work with and to understand their potential limitations and deficiencies.
It is \emph{comprehensive} in the sense that it provides a structured and critical description for each contained dataset with a level of detail not seen in other surveys before.
It is \emph{continuous} in the sense that we will continue adding further datasets in the future, extend existing entries, and potentially correct discovered errors.
Due to regular versioned releases with changelogs, users can directly track changes and reference fixed snapshots if desired.
\textsc{Comidds} is \emph{collaborative}, i.e., we strongly welcome contributions, both in the form of adding new dataset entries and improving existing ones.
At the moment, \textsc{Comidds} contains information on 48 datasets as well as various short paragraphs on related work (13 survey papers and nine websites).

\paragraph{Goals}

Motivated by the shortcomings of related surveys (cf. Sections~\ref{sec:introduction} and~\ref{sec:relatedwork}), we set the following goals for \textsc{Comidds}:
\begin{itemize}[leftmargin=*]
\item High coverage of datasets within our scope (see below): While the broadest survey that we found covers 52 datasets, we are striving to significantly exceed this number soon (cf. Section~\ref{sec:methodology}).
\item Actionable description: Each dataset should be represented in a way that researchers can profoundly decide which dataset(s) to use and how to interpret experimental results based on them.
\item Practical format: The survey should be easily accessible, extensible, maintainable, logically structured, and machine-readable.
\end{itemize}

\paragraph{Scope}

We currently focus on datasets suited for developing and evaluating methods for intrusion detection in \emph{enterprise networks}, i.e., environments usually involving client and server computers with common operating systems (particularly Windows and Linux), network hardware (e.g., routers, switches, firewalls), and typical  applications and services (e.g. web, mail, directory).
Adding datasets stemming from fundamentally different environments such as industrial control systems, Internet of Things, or otherwise specialized hardware or software is currently not planned by us, but might be considered if contributed by respective domain experts.

\paragraph{Features}

To begin with, all included datasets are summarized in an overview table, which comprises the following columns:
\begin{itemize}[leftmargin=*]
  \item the \textbf{name} of the dataset as introduced by its author(s),
  \item a very \textbf{brief description} of the dataset,
  \item the fundamental \textbf{data type(s)} contained: \textit{network} (e.g., network flows), \textit{host} (e.g., operating system log files), or \textit{both},
  \item the \textbf{year(s)} of creation or, if unknown, of publication,
  \item the basic \textbf{environment}, e.g., \emph{single system} or \emph{enterprise IT},
  \item the \textbf{operating system(s)}, e.g., \emph{Windows} or \emph{Linux},
  \item the \textbf{labeling}: \emph{direct} if data records are directly labeled as attack (class) or benign, \emph{indirect} if only indirect labeling such as periods of attack are given, and \emph{none} if no labels are present,
  \item the \textbf{data format(s)}, e.g., \emph{NetFlow}, \emph{syslog}, or \emph{Suricata alerts},
  \item the packed and unpacked \textbf{size} of the dataset in MB or GB.
\end{itemize}
In addition, for each dataset, there is a dedicated page containing an in-depth description, divided into the following sections:
\begin{itemize}[leftmargin=*]
  \item a \textbf{detailed table} showing concrete information beyond the summary table, such as \emph{attack categories} and \emph{benign activity},
  \item an \textbf{overview} summarizing the origin, purpose, and contents of the dataset in a few sentences,
  \item information on the \textbf{environment} in which the dataset was recorded, e.g., the involved systems and network architecture,
  \item what \textbf{activity} was performed while recording the dataset (either by humans or synthetically), both benign and adversarial, and
  \item which \textbf{files} are actually contained in the dataset (with respect to data sources, formats, and labeling);
  \item moreover a list of \textbf{relevant papers}, including the original publication of the dataset as well as independent analyses thereof,
  \item links to \textbf{relevant websites} (especially the download location),
  \item a list of \textbf{related datasets}, and finally
  \item \textbf{sample records} for each data format contained in the dataset (excluding binary formats such as pcap).
\end{itemize}

Appendix~\ref{sec:sample} exemplarily shows the description of the popular CSE-CIC-IDS2018 dataset~\cite{sharafaldin2018toward} as contained in the current version of \textsc{Comidds}.
Last but not least, all key characteristics can be downloaded as a CSV (comma-separated values) file to facilitate custom sorting, filtering, or plotting, as we will showcase in Section~\ref{sec:statistics}.

\begin{figure*}
  \centering
  \includegraphics[width=\linewidth]{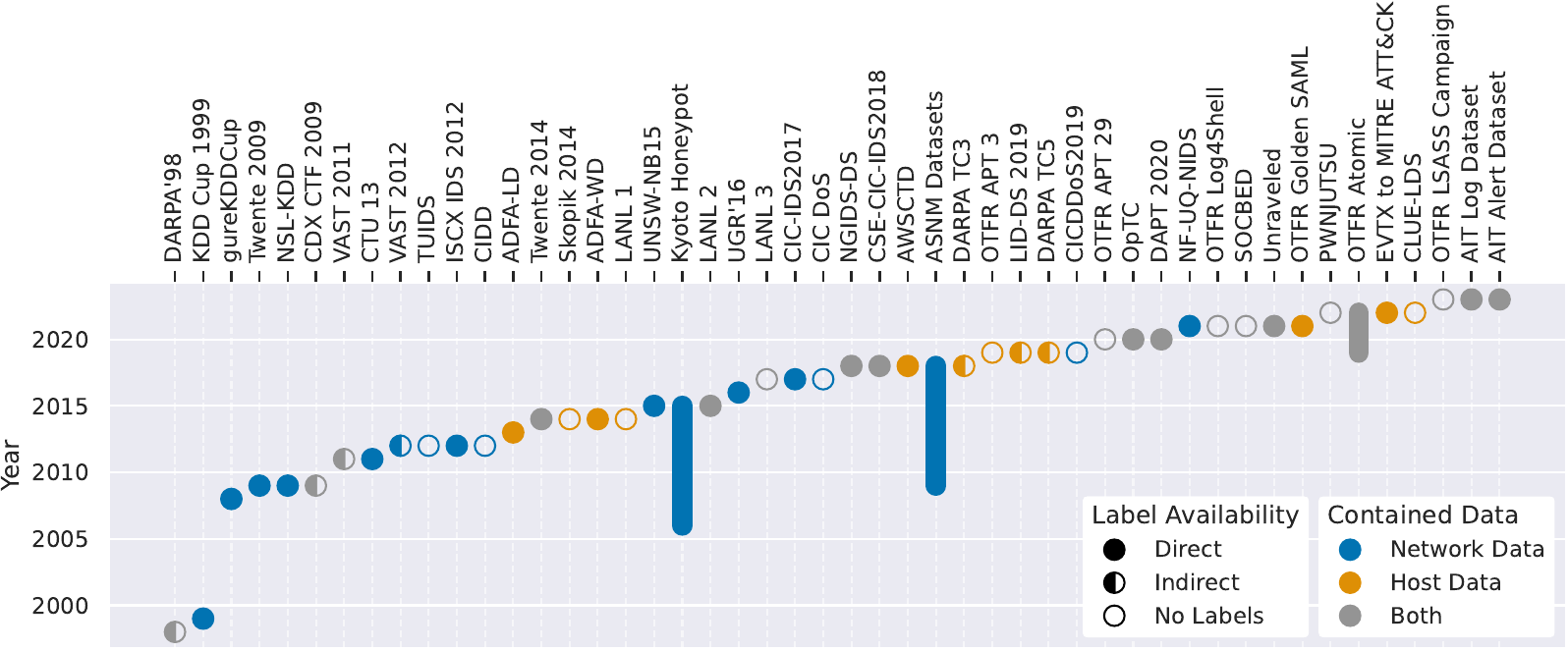}
  \caption{Age, labeling, and data types of all intrusion detection datasets surveyed until now (see Appendix~\ref{sec:list} for references)}
  \label{fig:datasets}
  \Description{A graph showing 48 dataset names on the x-axis and the years 1998 to 2023 on the y-axis. For each dataset, a symbol marks the year of creation or publication, the label availability (direct, indirect, or no labels), and the contained data (network data, host data, or both).}
\end{figure*}

\section{Survey Methodology}
\label{sec:methodology}

In the following, we describe our methodology for identifying relevant datasets in the literature and analyzing them, respectively.

\paragraph{Literature Review}

Prior to searching for original publications that contribute new intrusion detection datasets, we searched for already existing surveys of such datasets.
For this purpose, we leveraged Google Scholar combined with domain knowledge from personally known researchers working in this field.
We did not aim for a full coverage of such surveys since there is a large number of arguably redundant publications covering the same few datasets such as KDD Cup 1999~\cite{hettich1999kdd} or CSE-CIC-IDS2018~\cite{sharafaldin2018toward}, often discussing them with regards to some specific flaw or research question.
Instead, we focused on a selection of recent surveys offering the most comprehensive overview (cf. Section~\ref{sec:relatedwork}).

Using these surveys as a starting point, we found a total of 90 datasets that fit our scope.
To find further relevant datasets (especially those published after the latest surveys), we again utilized Google Scholar, using the search term \emph{``intrusion detection dataset''}, but limiting our search to works published in the year 2023 or later to keep the number of results manageable.
As this search resulted in 2030 works, we defined the following exclusion criteria:
\begin{itemize}[leftmargin=*]
\item The publication does not contribute its own novel dataset,
\item the contributed dataset is not publicly available,
\item the contributed dataset does not contain adversarial activity,
\item the publication is not available in English, or
\item the publication is not available in electronic form.
\end{itemize}

After applying these criteria, we were left with 30 publications that contribute their own dataset, of which only ten match our scope (i.e., a focus on enterprise networks).
Consequently, the total number of relevant datasets grew to exactly 100.

Lastly, we leveraged two more sources to find further datasets:
(1)~references within the selected publications (usually in the related work section) and (2)~the domain knowledge of researchers in this field, in both cases following the exclusion criteria as defined above.
This resulted in a grand total of 126 datasets. While this number might not be definitive, it excludes only those datasets that are not referenced by any major survey, not cited in any of these 126 works, and are unknown to several domain experts.
At the time of writing, \textsc{Comidds} already covers 48 of these 126 datasets, focusing on the most popular ones, with more being added continuously. 
To include datasets into \textsc{Comidds}, we analyze them as follows.

\paragraph{Dataset Analysis}

There are a number of dataset characteristics that various researchers regard as desirable, e.g., \emph{documentation of labeling methodology}~\cite{landauer2023maintainable,uetz2021socbed,sharafaldin2018toward}.
In an ideal world, every dataset would fulfill all of these characteristics, while also describing the process leading to their fulfillment.
In reality, few publications document such issues, making it difficult to determine whether or not characteristics are present/fulfilled.
For example, many works describe the simulation of benign activity in just a few sentences, making it close to impossible to determine if or to which extent the requirement of \emph{realistic benign activity} is fulfilled.

Consequently, we do not aim to check each dataset against all requirements proposed in the literature, both because it is not feasible and requirements are often vague, making classification difficult or sometimes impossible, even with a lot of effort.
Instead, we resort to an approach in part similar to that of Ring et al.~\cite{ring2019survey}, defining key characteristics (cf. Section~\ref{sec:survey}) and reviewing all datasets with respect to them.
We believe that this information serves as a sufficient representation for a given dataset, providing researchers with the means to quickly obtain detailed information and decide if this dataset could be suitable for their current undertaking.
At the same time, this level of detail allows us to spend a feasible amount of time per dataset (usually a few hours).

During our analysis, we found that 23 of the 126 identified datasets are not backed by an academic publication or otherwise sufficient documentation.
We decided that these datasets do not undergo the full analysis process as described above, but are instead listed separately on the \textsc{Comidds} website and each described in a single paragraph since at least some of the key characteristics cannot be determined from the documentation.
However, we found cases where well-defined parts of such works were documented in a dedicated paper, thus being eligible for the previously described analysis process.
For example, the \emph{Malware Capture Facility Project}~\cite{garcia2014mcfp} provides a large number of pcap files collected from real networks, with little to no explanation for each of them -- except for a select subset, known as CTU-13, which has its own paper~\cite{garcia2014botnet} and is thus included in \textsc{Comidds}.
We continue our discussion with presenting statistics of the datasets analyzed so far.

\section{Datasets Statistics}
\label{sec:statistics}

Since \textsc{Comidds} provides key characteristics of all surveyed datasets in machine-readable CSV format, generating statistics and plotting them is straightforward.
To illustrate this, we present two exemplary visualizations created from the data in the CSV file.
They are also available in the repository (including source code) and updated automatically whenever dataset entries are added or changed.

Figure~\ref{fig:datasets} depicts all datasets surveyed so far, where the y-axis shows the year of data creation or, if unknown, of publication.
Datasets comprising more than one year are visualized accordingly.
In addition, data types and label availability are shown (cf. Section~\ref{sec:survey}).
Note that while this figure provides a broad overview of the current datasets landscape, it also simplifies some aspects.
For example, while the DARPA’98 and CSE-CIC-IDS2018 datasets contain both network and host data and are visualized as such, only their network data is labeled and thus typically used by other researchers.

\begin{figure}
  \centering
  \includegraphics[width=\linewidth]{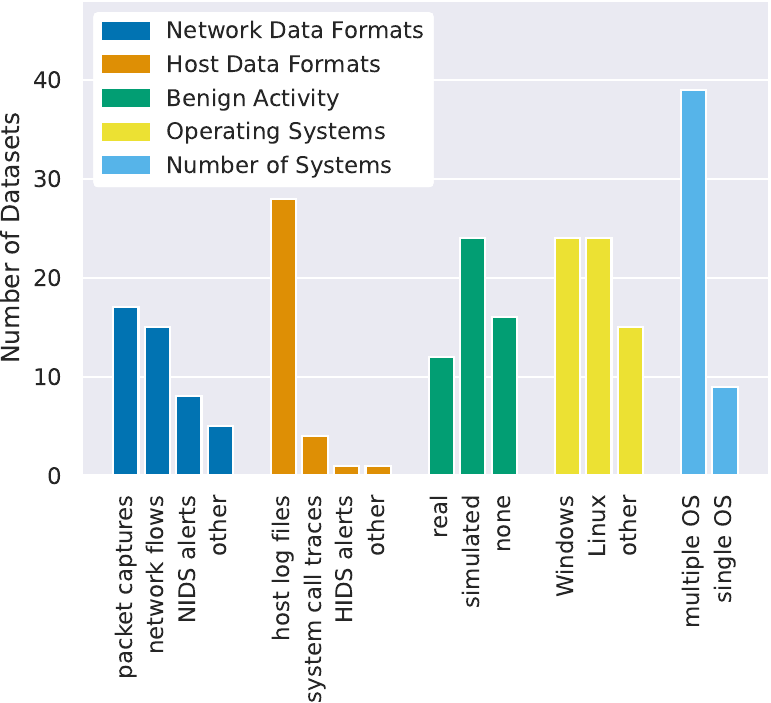}
  \caption{Characteristics of the surveyed datasets}
  \label{fig:datatypes}
  \Description{A bar plot showing the number of surveyed datasets that fulfill various characteristics with respect to data formats, benign activity, and involved operating systems.}
\end{figure}

Figure~\ref{fig:datatypes} plots multiple characteristics of the surveyed datasets, grouped into five categories: Network data formats, host data formats, type of benign activity, involved operating systems, and number of systems (in the sense of data-generating operating systems).
Except for the last category, these classifications are not mutually exclusive, so the sum of a category does not necessarily match the total number of datasets surveyed.
Note that we deliberately omit some characteristics, namely, dataset size, runtime, and number of machines, since we generally find them ineligible for a qualitative comparison of datasets with one another.
For example, datasets containing packet captures can be orders of magnitude larger than NetFlow datasets despite resembling less activity.
Similarly, runtime and number of machines do not necessarily correlate with the quantity and quality of benign or adversarial activity.

\section{Related Work}
\label{sec:relatedwork}

While there exists a multitude of publications touching upon the topic of intrusion detection datasets, our discussion of related work focuses on works that share our principal goal of providing a broad yet actionable overview to help researchers choose appropriate datasets and understand their respective limitations as well as potential deficiencies.

Gümüşbaş et al.~\cite{gumucsbacs2020survey} discuss various intrusion detection methods based on deep learning, alongside which they present a list of datasets commonly used to benchmark these approaches.
Twenty network-based datasets are described in a short manner, with the six most frequently cited undergoing further analysis regarding properties such as number of features and attack types.
Bridges et al.~\cite{bridges2019hidssurvey} provide a survey focused on methods and datasets leveraging host data.
They offer an overview of 22 datasets in the form of a brief description for each, listing information such as origin or data types, though not in a consistent manner.
Yang et al.~\cite{yang2022anomalysurvey} compiled the broadest survey listed here, covering a large variety of publications and topics related to anomaly-based network intrusion detection, ranging from data preprocessing over evaluation metrics to datasets.
They cover 52 datasets, although very little detail is provided for each of them.

Other surveys place their emphasis solely on datasets themselves.
Ring et al.~\cite{ring2019survey} provide the most in-depth overview of all studied papers, doing so by first defining 15 different dataset properties to describe, such as year of creation, format, duration, or type of network, and then applying this methodology to 34 network datasets, along with a description.
Kenyon et al.~\cite{kenyon2020fit} follow a similar approach, supplying a short paragraph per dataset but defining fewer features (origin, anonymization, data types, attack types).
Furthermore, they define characteristics a dataset should fulfill in order to be suitable for intrusion detection research, and discuss a selection of the datasets with respect to them.

Lastly, in addition to these surveys, there are several works featuring a substantially smaller number of datasets, with the goal of answering specific research questions.
For example, Landauer et al.~\cite{landauer2023critical} and Engelen et al.~\cite{engelen2022pillars} analyze six and five popular datasets, respectively, discussing flaws affecting anomaly-based approaches and their consequences on state-of-the-art research.
However, as the objective of our work is to offer a comprehensive survey that helps researchers to narrow down dataset choices, we consider their work to be complementary to ours.

A unifying property of the discussed broad surveys~\cite{gumucsbacs2020survey,bridges2019hidssurvey,yang2022anomalysurvey,ring2019survey,kenyon2020fit}, as well as a driving motivator for the creation of \textsc{Comidds}, is their lack of detail required to choose an appropriate dataset and become aware of its potential limitations and deficiencies.
While certainly helpful in providing a general overview of a portion of currently existing intrusion detection datasets, researchers looking to process a dataset for their specific use case will have to invest substantial amounts of time into manual analysis, or resort to one of the most popular datasets (e.g., CSE-CIC-IDS2018~\cite{sharafaldin2018toward}) without questioning its suitability.
As an example, only Ring et al.~\cite{ring2019survey} provide basic information such as the data format (and even then only differentiate between ``packet'', ``flow'', and ``other'') and none provide actionable information on the simulation environment, ongoing activity within that environment, or samples of contained data -- all of which are highly relevant, if not crucial, for performing and evaluating experiments based on these datasets.

\section{Conclusion}

This work addresses the challenges of selecting suitable datasets for intrusion detection research while taking into account their characteristics and potential deficiencies.
We found that existing dataset surveys have significant shortcomings in the sense that they are static and either incomprehensive or superficial.
With \textsc{Comidds}, we strive to resolve these issues by providing a repository-based survey that is comprehensive, continuous, and collaborative.
\textsc{Comidds} currently covers 48 datasets and allows for sorting, filtering, and plotting of key characteristics to facilitate dataset selection.
We will regularly add new datasets in the future and welcome contributions from other researchers.
In addition, we intend to add further automatically-updated statistics and plots to the website.
Ultimately, we hope that \textsc{Comidds} gains acceptance as a reference survey for intrusion detection datasets within its scope and thereby facilitates sound research in this practically relevant field.

\begin{acks}
We would like to thank Frédéric Majorczyk and Maxime Lanvin for providing feedback on \textsc{Comidds} and suggesting additional datasets and papers.
We also thank the anonymous reviewers for their time and valuable comments on the paper.
\end{acks}

\bibliographystyle{ACM-Reference-Format}
\bibliography{references}

\appendix

\section{Exemplary Dataset Entry}
\label{sec:sample}

This section shows information on the popular CSE-CIC-IDS2018 dataset as contained in \textsc{Comidds} as a concrete example for one of the currently 48 covered datasets.
Note that this entry might be extended, corrected, or otherwise improved in future releases.

\begin{table}
  \caption{Detailed table for the exemplary CSE-CIC-IDS2018 dataset as contained in the current version of \textsc{Comidds}}
  \label{tab:csecicids2018}
  \begin{tabular}{ll}
    \toprule
    \textbf{Network Data Source(s)} & pcaps, NetFlows\\
    \textbf{Network Data Labeled} & Yes, NetFlows are labeled\\
    \textbf{Host Data Source(s)} & Ubuntu \& Windows event logs\\
    \textbf{Host Data Labeled} & No\\
    \midrule
    \textbf{Overall Setting} & Enterprise IT\\
    \textbf{OS Types} & Windows 7/8/10/Vista/Server 2016,\\
      & Ubuntu 14.04/16.04, MacOS;\\
      & Kali \& Windows 8.1 (Attacker)\\
    \textbf{Number of Machines} & 450\\
    \textbf{Total Runtime} & \textasciitilde{}5 days\\
    \textbf{Year of Collection} & 2018\\
    \textbf{Attack Categories} & Bruteforce, Heartbleed, Botnet,\\
      & DoS/DDoS, Web-Based, Infiltration\\
      & from Inside Network\\
    \textbf{Benign Activity} & Synthetic, models complex behavior\\
    \midrule
    \textbf{Packed Size} & 220\,GB\\
    \textbf{Unpacked Size} & n/a\\
    \textbf{Download Link} & Instructions at bottom of page\\
    \bottomrule
  \end{tabular}
\end{table}

\subsection*{Overview}

A collaboration between the Communications Security Establishment (CSE) and the Canadian Institute for Cybersecurity (CIC), this dataset uses the notion of profiles to generate cybersecurity datasets in a systematic manner, including various attack types and a large and diverse infrastructure.
It is a continuation of previous efforts (CIC IDS2017), featuring similar attacks and benign behavior, but being significantly larger in scale (14 vs. 450 victim machines, 1~vs. 6 victim networks).
While being one of the primary benchmark datasets in the current field of NIDS research, researchers have discovered errors within this dataset, affecting aspects like attack orchestration, feature generation, or labeling.
Essential details of this dataset are summarized in Table~\ref{tab:csecicids2018}.

\subsection*{Environment}

The attacking infrastructure contains 50 machines, the victim infrastructure consists of 5 departments with a total of 420 PCs and 30 servers.
An overview is provided by the diagram below [note: omitted in this paper to save space].
Presumably, vulnerable software versions have been installed to facilitate certain exploits, but this is more suggested than specified in their description.

\subsection*{Activity}

Simulated behavior is defined in the form of profiles, divided into benign (B) and malicious (M) profiles.
B-profiles are derived from observing human behavior, from which some features are learned/ extracted.
M-profiles consist of seven different attack scenarios, each based on a certain attack type: Bruteforce, Heartbleed, Botnet, DoS, DDoS, Web-Based, and Infiltration from Inside Network.
The total capturing period lasted \textasciitilde{}5 days, with attacks being performed on every day except the first.
Details for each attack as well as the timing are available on the linked homepage.

\subsection*{Contained Data}

The dataset includes the network traffic and log files of each victim machine, combined with 80 network features extracted from captured traffic using \href{https://www.unb.ca/cic/research/applications.html#CICFlowMeter}{CICFlowMeter}.
Data is divided into two main directories, \texttt{Network Traffic and Log Data} as well as \texttt{Processed Traffic Data for ML Algorithms}, with data being organized per day, respectively.
The former contains raw data in the form of unlabeled network traffic (pcap) and event logs (Windows/Ubuntu).
The latter consists of labeled features derived from the aforementioned network traffic (although the labeling logic is not transparently documented);
these features are what is most commonly leveraged when using this dataset.
Each feature is explained in detail on the homepage linked below.
The aforementioned flaws of this dataset, such as some simulation artifacts making detection artificially easy, are for example laid out in Paper~2 referenced below.

\subsection*{Example Data}

Note: We only show a small excerpt of the example data in this paper to give an idea of the structure on the \textsc{Comidds} website.

\subsubsection*{Labeled features from ``Processed Traffic Data for ML Algorithms/\\Thursday-01-03-2018\_TrafficForML\_CICFlowMeter.csv'':}

\begin{verbatim}
Dst Port,Protocol,Timestamp,Flow Duration,Tot Fwd...
3389,6,01/03/2018 09:56:59,4046191,14,7,1386,392,...
58655,6,01/03/2018 09:56:59,86620951,2,0,0,0,0,0,...
50657,6,01/03/2018 09:56:59,0,2,0,0,0,0,0,0,0,0,0,...
\end{verbatim}

\subsubsection*{Ubuntu event logs taken from ``Network Traffic and Log data/\\Friday-16-02-2018/logs/U172.31.69.25'':}

\begin{verbatim}
Feb 16 07:39:01 ip-172-31-69-25 CRON[11625]: (root)...
Feb 16 07:48:09 ip-172-31-69-25 dhclient[922]: ...
Feb 16 07:48:09 ip-172-31-69-25 dhclient[922]: ...
\end{verbatim}

\subsection*{Papers}

\begin{itemize}[leftmargin=*]
\item[1] \href{https://doi.org/10.5220/0006639801080116}{Toward Generating a New Intrusion Detection Dataset and Intrusion Detection Traffic Characterization (2017)}
\item[2] \href{https://doi.org/10.1109/CNS56114.2022.9947235}{Error Prevalence in NIDS datasets: A Case Study on CIC-IDS-2017 and CSE-CIC-IDS-2018 (2022)}
\end{itemize}

\subsection*{Links}

\begin{itemize}[leftmargin=*]
\item \href{https://www.unb.ca/cic/datasets/ids-2018.html}{Homepage}. For download, install AWS CLI and run \texttt{aws s3 sync --no-sign-request --region <your-region> "s3://cse-cic-\\ids2018/" dest-dir}, where your-region is your AWS region and destination-dir is the target directory. If you only need the labeled features, use \texttt{s3://cse-cic-ids2018/Processed Traffic Data for ML Algorithms} as your URL.
\item \href{https://registry.opendata.aws/cse-cic-ids2018/}{Secondary Source}
\end{itemize}

\subsection*{Related Entries}

\begin{itemize}[leftmargin=*]
\item CIC IDS2017
\item NF-UQ-NIDS
\end{itemize}

\section{List of Covered Datasets}
\label{sec:list}

The following intrusion detection datasets are currently described in \textsc{Comidds}, ordered by year of creation/publication (newest first):

\begin{enumerate}[topsep=3pt]
\item AIT Alert Dataset \cite{landauer2023introducing}
\item AIT Log Dataset \cite{landauer2023maintainable}
\item OTFR Security Datasets - LSASS Campaign \cite{rodriguez2024security}
\item CLUE-LDS \cite{landauer2022clue}
\item EVTX to MITRE ATT\&CK \cite{crevoisier2024evtx}
\item OTFR Security Datasets - Atomic \cite{rodriguez2024security}
\item PWNJUTSU \cite{berady2022pwnjutsu}
\item NF-UQ-NIDS \cite{sarhan2022towards}
\item OTFR Security Datasets - Log4Shell \cite{rodriguez2024security}
\item OTFR Security Datasets - SimuLand Golden SAML \cite{rodriguez2024security}
\item SOCBED Example Dataset \cite{uetz2021socbed}
\item Unraveled \cite{myneni2023unraveled}
\item DAPT 2020 \cite{myneni2020dapt}
\item OpTC \cite{darpa2020optc}
\item OTFR Security Datasets - APT 29 \cite{rodriguez2024security}
\item CICDDoS2019 \cite{sharafaldin2019ddos}
\item DARPA TC5 \cite{darpa2020tc}
\item LID-DS 2019 \cite{grimmer2019modern}
\item OTFR Security Datasets - APT 3 \cite{rodriguez2024security}
\item ASNM Datasets \cite{homoliak2020asnm}
\item AWSCTD \cite{ceponis2018towards}
\item CSE-CIC-IDS2018 \cite{sharafaldin2018toward}
\item DARPA TC3 \cite{darpa2020tc}
\item NGIDS-DS \cite{waqas2018developing}
\item CIC DoS \cite{hadian2017detecting}
\item CIC-IDS2017 \cite{sharafaldin2018toward}
\item Unified Host and Network Data Set \cite{turcotte2018unified}
\item UGR’16 \cite{maciafernandez2018ugr}
\item Comprehensive, Multi-Source Cyber-Security Events \cite{kent2016cyber}
\item Kyoto Honeypot \cite{song2011kyoto}
\item UNSW-NB15 \cite{moustafa2015unsw}
\item ADFA-WD \cite{creech2014afda}
\item Skopik 2014 \cite{skopik2014semi}
\item Twente 2014 \cite{hofstede2014ccr}
\item User-Computer Associations in Time \cite{hagberg2014connected,kent2014authdata}
\item ADFA-LD \cite{creech2014semantic,creech2013generation,creech2014afda}
\item CIDD \cite{kholidy2012cidd}
\item ISCX IDS 2012 \cite{shiravi2012toward}
\item TUIDS \cite{gogoi2012packet}
\item VAST Challenge 2012 \cite{cook2012vast}
\item CTU 13 \cite{garcia2014botnet}
\item VAST Challenge 2011 \cite{vast2011}
\item CDX CTF 2009 \cite{sangster2009toward}
\item NSL-KDD \cite{tavallaee2009kdd}
\item Twente 2009 \cite{sperotto2009labeled}
\item gureKDDCup \cite{perona2017generation}
\item KDD Cup 1999 \cite{hettich1999kdd}
\item DARPA’98 Intrusion Detection Program \cite{lippmann2000darpa}
\end{enumerate}

\noindent We will continue adding and improving dataset entries in the future.

\end{document}